\documentclass[aps,prl,twocolumn,superscriptaddress,showpacs]{revtex4}
\usepackage{graphicx}

\begin{document}

\title{Quasiparticle Corrections to the Electronic Properties of Anion Vacancies at GaAs(110) and InP(110)}
\author{Magnus Hedstr\"om}
\affiliation{Fritz-Haber-Institut der Max-Planck-Gesellschaft, Faradayweg 4--6, 14195 Berlin-Dahlem, Germany}
\author{Arno Schindlmayr}
\email{A.Schindlmayr@fz-juelich.de}
\affiliation{Fritz-Haber-Institut der Max-Planck-Gesellschaft, Faradayweg 4--6, 14195 Berlin-Dahlem, Germany}
\affiliation{Institut f\"ur Festk\"orperforschung, Forschungszentrum J\"ulich, 52425 J\"ulich, Germany}
\author{G\"unther Schwarz}
\altaffiliation[Present address: ]{Lehrstuhl f\"ur Theoretische Festk\"orper\-physik, Universit\"at Erlangen-N\"urnberg, Staudtstra{\ss}e~7, 91058 Erlangen, Germany}
\author{Matthias Scheffler}
\affiliation{Fritz-Haber-Institut der Max-Planck-Gesellschaft, Faradayweg 4--6, 14195 Berlin-Dahlem, Germany}
\date{\today}

\begin{abstract}
We propose a new method for calculating optical defect levels and thermodynamic charge-transition levels of point defects in semiconductors, which includes quasiparticle corrections to the Kohn-Sham eigenvalues of density-functional theory. Its applicability is demonstrated for anion vacancies at the (110) surfaces of III-V semiconductors. We find the (+/0) charge-transition level to be 0.49 eV above the surface valence-band maximum for GaAs(110) and 0.82 eV for InP(110). The results show a clear improvement over the local-density approximation and agree closely with an experimental analysis.
\end{abstract}


\maketitle

The electrical and optical properties of semiconductors depend sensitively on the electronic structure in the gap region and can hence be modified dramatically by the presence of native defects and impurities that introduce unwanted additional states inside the fundamental band gap. Most importantly, such electrically active defects can trap charge carriers (electrons or holes), counteracting the effect of intentional doping. If their concentration is sufficiently high, this process can lead to a full compensation of implanted acceptors and donors and thus eventually to Fermi-level pinning. Besides, electron-hole recombination at deep defects drastically reduces the lifetime of minority carriers, and transitions involving defect states inside the band gap may dominate optical absorption. This is especially relevant at surfaces and interfaces, where the crystal termination and the contact with other phases naturally give rise to a high number of structural defects that have been linked to the formation of Schott\-ky barriers \cite{spicer80}. The central quantities are the optical defect levels inside the band gap and the charge-transition levels. The former can, in principle, be probed by direct (filled states) or inverse (empty states) photoemission. The Franck-Condon principle is well justified, as the rearrangement of the atoms happens on a much slower time scale than the electron emission or absorption, but the coupling to the atomic lattice may be visible in the line widths and shapes. The optical defect levels contain the full \emph{electronic} relaxation in response to the created hole or the injected electron, however. The charge-transition levels, on the other hand, are thermodynamic quantities and specify the values of the Fermi energy where the charge state of the defect changes. Therefore, they are affected noticeably by the atomic relaxation taking place upon the addition or removal of an electron.

Despite considerable efforts, a reliable determination of the optical defect levels and the related charge-transition levels of deep defects still poses a very difficult challenge. Experimental investigations are thwarted by the fact that many traditional spectroscopic techniques are not applicable due to the low concentration of native point defects, while capacitance methods like deep-level transient spectroscopy \cite{lang74} are very sensitive but provide no elemental or structural information to identify the type of defect. For surfaces, at least, Ebert \textit{et al.}\ \cite{ebert00} now demonstrated how the electronic structure of individual defects can be deduced using a combination of scanning tunneling microscopy (STM) and photoelectron spectroscopy, providing the first reliable experimental analysis of a charge-transition level for the P vacancy at InP(110). Theoretical approaches, on the other hand, must include accurate exchange-correlation contributions, the coupling between electronic and lattice degrees of freedom, and, in general, require the treatment of open systems in which the number of particles is not constant \cite{scherz93}. Previous studies that employed density-functional theory (DFT) in the local-density approximation (LDA) \cite{hohenberg64} indeed suffered from fundamental limitations \cite{ebert00,zhang96,kim96,qian02}. For example, Ebert \textit{et al.}\ \cite{ebert00} noted ``that the systematic error for the calculated energies of the charge transfer levels is too large to identify the symmetry of the vacancy on the position of the defect level only.'' In order to overcome this problem we here propose a new computational approach, broadly applicable to defects in the bulk as well as at surfaces, that combines DFT with many-body perturbation theory. As an example, we examine the optical defect levels and the thermodynamic charge-transition levels of anion vacancies at GaAs(110) and InP(110). The results are in close agreement with the experimental analysis \cite{ebert00}.

The geometry of anion vacancies at the (110) surfaces of III-V semiconductors is now well understood thanks to a combination of experimental and theoretical studies. For $p$-doped materials STM images of the filled states under negative bias feature a localized hole at the position of the missing anion surrounded by a voltage-dependent depression \cite{lengel94}. The latter is due to a downward local band bending, indicating a positive charge of the vacancy. The charge state is, in fact, established as +1 \cite{chao96}, which is also predicted by electronic-structure calculations \cite{ebert00,zhang96,kim96,qian02}. STM images acquired under positive bias probe the empty $p_z$-like orbitals of the cation sublattice and show an enhancement of the cations surrounding the vacancy, initially wrongly interpreted as an upward relaxation of those atoms \cite{lengel94} but now understood as arising from the local depression of the electron density. DFT-LDA calculations actually show that an \emph{inward} relaxation of the two Ga atoms enclosing the As vacancy $V_\mathrm{As}^+$ at GaAs(110) is consistent with the observations \cite{zhang96,kim96}. The symmetry of the positively charged anion vacancies was initially a matter of controversy \cite{zhang96,kim96}. In STM images they appear symmetric, but in a combined experimental and theoretical study of $V_\mathrm{P}^+$ at InP(110) Ebert \textit{et al.}\ \cite{ebert00} explained the observed features as resulting from the thermal flip motion between two degenerate asymmetric configurations. This interpretation later received further confirmation \cite{qian03}. In $n$-doped materials, on the other hand, the vacancy is in a charge state of $-1$ \cite{domke98}, and DFT-LDA calculations predict a symmetric relaxation for this configuration as well as the neutral vacancy \cite{zhang96,kim96}.

The anion vacancies at GaAs(110) or InP(110) give rise to three nondegenerate electronic states, labeled 1$a'$, 1$a''$, and 2$a'$. While the 1$a'$ state is located several eV below the valence-band maximum and always filled with two electrons, and the 2$a'$ state is too high in energy to become populated, the 1$a''$ state lies inside the band gap. It is this state, therefore, that is relevant for the discussion of charge-transition levels. Depending on the level of doping it may be occupied by zero, one, or two electrons, which corresponds to the positive, neutral, and negative charge state, respectively.

The formation energy of a surface vacancy with charge state $q$, relative to that of the neutral defect, is given by
\begin{equation}
E^\mathrm{form}(q/0) = E^\mathrm{vac}(q,Q_q) - E^\mathrm{vac}(0,Q_0) + q \epsilon_\mathrm{F}\;,
\end{equation}
where $E^\mathrm{vac}(q,Q)$ denotes the total energy of a surface featuring a single vacancy with the actual electron population $q \in \{+,0,-\}$ and geometry optimized for charge state $Q \in \{Q_+,Q_0,Q_-\}$. The final term accounts for the transfer of the charge $q$ between the defect level and the electron reservoir, i.e., the Fermi energy $\epsilon_\mathrm{F}$. The charge-transition levels $\epsilon^{q/q'}$ are defined as the values of $\epsilon_\mathrm{F}$ where the charge state of the vacancy changes, i.e., where $E^\mathrm{form}(q/0) = E^\mathrm{form}(q'/0)$, and conventionally given relative to the surface valence-band maximum. For the systems considered here the interesting transitions are $\epsilon^{+/0}$ and $\epsilon^{0/-}$. For example, the former is
\begin{equation}\label{Eq:ctlevel}
\epsilon^{+/0} = E^\mathrm{vac}(0,Q_0) - E^\mathrm{vac}(+,Q_+)\;.
\end{equation}
All previous DFT-LDA calculations for surface point defects evaluated this energy difference directly \cite{ebert00,zhang96,kim96,qian02}, but this approach leads to systematic errors that arise because the total energies $E^\mathrm{vac}(0,Q_0)$ and $E^\mathrm{vac}(+,Q_+)$ refer to systems with different electron numbers. As is well known, the exact exchange-correlation potential in DFT exhibits a discontinuity upon addition or removal of an electron \cite{perdew83}, which is not contained in the LDA or other jellium-based functionals. Besides, aspects like the self-interaction are treated inappropriately. As a consequence, the band gaps of semiconductors and the energies of localized defect states are not given correctly: for the P vacancy at InP(110) the experimentally determined $\epsilon^{+/0}$ level of 0.75$\pm$0.1 eV \cite{ebert00} contrasts noticeably with the calculated values 0.52 eV \cite{ebert00} and 0.388 eV \cite{qian02}. The variation between the two theoretical results can be traced to differences in the pseudopotentials and parallels the variation of the corresponding band gaps.

To arrive at a more accurate quantitative method that corrects the above-mentioned severe shortcomings of the LDA we rewrite Eq.\ (\ref{Eq:ctlevel}) as
\begin{eqnarray}\label{Eq:ctlevel1}
\epsilon^{+/0} &=& \left[ E^\mathrm{vac}(+,Q_0) - E^\mathrm{vac}(+,Q_+) \right]\\ &&+ \left[ E^\mathrm{vac}(0,Q_0) - E^\mathrm{vac}(+,Q_0) \right]\nonumber
\end{eqnarray}
by adding and subtracting the total energy $E^\mathrm{vac}(+,Q_0)$ of a system with the geometry of the relaxed neutral vacancy but a charge state $q = +1$. In this way the charge-transition level is decomposed into two separate contributions. The first describes the structural relaxation energy for the positive charge state and two different geometries: that of the neutral and that of the positive charge state. It is always positive. As the electron number remains constant, the problem of the discontinuity does not arise, and DFT-LDA is perfectly applicable. The second term equals the ionization energy of the neutral defect, where the removed electron is transfered to the reservoir, i.e., the Fermi energy. Determining the ionization potential from the quasiparticle band structure requires a correction of the Kohn-Sham eigenvalues, for which we employ many-body perturbation theory. Specifically, we use the $G_0W_0$ approximation for the electronic self-energy \cite{hedin65}. This approach is known to yield reliable band gaps for III-V semiconductors \cite{godby87} and their surfaces \cite{zhu89}.

In the same spirit, $\epsilon^{0/-}$ can be written as
\begin{eqnarray}\label{Eq:ctlevel2}
\epsilon^{0/-} &=& \left[ E^\mathrm{vac}(-,Q_-) - E^\mathrm{vac}(-,Q_0) \right]\\ &&+ \left[ E^\mathrm{vac}(-,Q_0) - E^\mathrm{vac}(0,Q_0) \right]\nonumber\;.
\end{eqnarray}
The first term in this case describes the energy difference of the vacancy with $q = -1$ between its own equilibrium geometry and that of the neutral charge state. It is always negative. The second term equals the electron affinity of the neutral charge state. In principle, the self-energy of the neutral charge state yields both the ionization potential and the electron affinity, which correspond to the energy of the highest occupied and the lowest unoccupied quasiparticle state, respectively. From a computational point of view, however, this procedure is inconvenient, because the neutral defect has an odd number of electrons and requires a spin-polarized calculation. Instead, we follow an equivalent approach and extract the energy levels from two separate calculations for non-spin-polarized systems with an even number of electrons. In practice, we thus determine $E^\mathrm{vac}(0,Q_0) - E^\mathrm{vac}(+,Q_0)$ as the electron affinity of the positive charge state and $E^\mathrm{vac}(-,Q_0) - E^\mathrm{vac}(0,Q_0)$ as the ionization potential of the negative charge state, both in the $Q_0$ geometry.

In the following we apply the expressions derived above to anion vacancies at the (110) surfaces of GaAs and InP. To determine the defect geometries we use DFT together with norm-conserving pseudopotentials \cite{bockstedte97} and the LDA exchange-correlation functional \cite{ceperley80}. The surfaces are simulated using a supercell with a (2$\times$4) periodicity in the [001] and [1\=10] directions, consisting of six atomic layers separated by a vacuum buffer equivalent to four layers. A single vacancy is created at one side of the slab, while the dangling bonds at the other are passivated by pseudoatoms with noninteger nuclear charges of 0.75 and 1.25 for anion and cation termination, respectively. This mimics the continuation of the substrate by a III-V bulk layer. We use the theoretical lattice constants 5.55 \AA\ for GaAs and 5.81 \AA\ for InP to prevent errors resulting from a nonequilibrium unit-cell volume during the surface relaxation. The integration in reciprocal space is carried out with a mesh corresponding to eight $\mathbf{k}$-points in the two-dimensional Brillouin zone of the (1$\times$1) unit cell of the defect-free surface. In the case of charged defects we apply a uniform compensating background in order to ensure overall charge neutrality. For GaAs we thus obtain the relaxation energies $E^\mathrm{vac}(+,Q_0) - E^\mathrm{vac}(+,Q_+) = 0.30$ eV and $E^\mathrm{vac}(-,Q_-) - E^\mathrm{vac}(-,Q_0) = -0.13$ eV. The corresponding values for InP are 0.20 eV and $-0.17$ eV.

\begin{figure}
\centerline{\includegraphics[width=\columnwidth,clip]{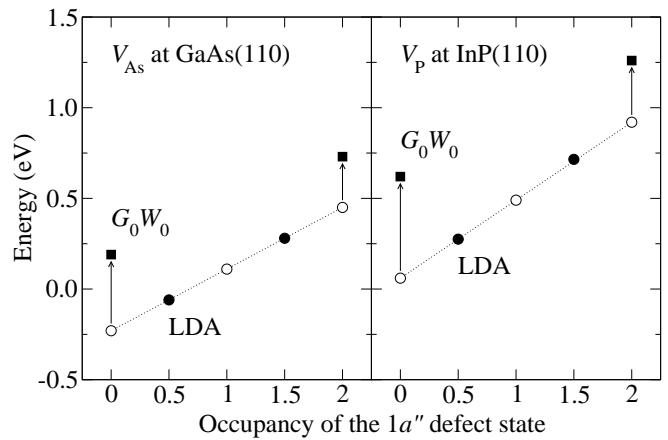}}
\caption{Position of the optical 1$a''$ defect level of the anion vacancies at GaAs(110) and InP(110) as a function of the occupation number. The geometry is identical in all calculations for the same material and optimized for the neutral charge state. The Slater-Janak transition-state approach (filled circles) yields the ionization potentials and electron affinities in the LDA\@. More accurate results are obtained by calculating the quasiparticle corrections within the $G_0W_0$ approximation (filled squares). In either case the energy zero is set to the respective surface valence-band maximum.}\label{Fig:variation}
\end{figure}

Before presenting our quasiparticle results, we first calculate the charge-transition levels strictly within the LDA by invoking the Slater-Janak transition-state approach \cite{slater72}, where the ionization potential and the electron affinity equal the eigenvalue of the 1$a''$ level determined self-consistently with the noninteger occupancy 0.5 and 1.5, respectively. The transition state corrects, at least partially, the erroneous self-interaction of the LDA but not the discontinuity problem. The resulting energy contributions are displayed in Fig.~\ref{Fig:variation} relative to the surface valence-band maximum. The occurrence of slightly negative values in some cases implies that the defect level actually falls below the valence-band maximum. This is an artefact of the constrained nonequilibrium geometry: if the atomic structure is allowed to relax, then the defect level always lies inside the band gap. For all systems studied here we find that the transition-state approach yields the same results, with a deviation of less than 0.01 eV, as a straightforward evaluation of Eq.\ (\ref{Eq:ctlevel}) and the corresponding formula for $\epsilon^{0/-}$ within the LDA\@. Quantitatively, our calculated value of 0.47 eV for $\epsilon^{+/0}$ in InP exhibits the same systematic underestimation of the experimentally derived charge-transition level as earlier studies at this level of approximation \cite{ebert00,qian02}.

In order to determine the electronic contribution to the charge-transition levels more accurately we employ many-body perturbation theory. The energies derived within this framework correspond directly to the values measured in direct or inverse photoemission. We follow the usual approach to calculate the quasiparticle energies
\begin{equation}
\epsilon_{1a''} = \epsilon_{1a''}^\mathrm{KS} + \langle \varphi_{1a''}^\mathrm{KS} | \Sigma(\epsilon_{1a''}) - V_\mathrm{xc} | \varphi_{1a''}^\mathrm{KS} \rangle
\end{equation}
as a first-order correction of the Kohn-Sham eigenvalues $\epsilon_{1a''}^\mathrm{KS}$. Here $\Sigma$ is the complex, nonlocal, and frequency-dependent self-energy, which we evaluate in the $G_0W_0$ approximation using the Green function $G_0$ of the underlying Kohn-Sham system. Our numerical implementation is based on the space-time method \cite{rieger99}. The local exchange-correlation potential $V_\mathrm{xc}$ must be subtracted from the self-energy to avoid double counting. A more detailed discussion of our computational method can be found in Ref.\ \cite{hedstrom02}. The quasiparticle corrections, but not the Kohn-Sham eigenvalues, are obtained from a smaller (2$\times$2) surface cell, which reduces the computational effort considerably. Although we investigate charged systems, the relative self-energy shifts are insensitive to the size of the supercell, because they only include exchange-correlation effects and no electrostatic Hartree contribution. Keeping the atomic positions fixed at the optimized geometry for the neutral vacancy, we performed separate $G_0W_0$ calculations for the positive and negative charge states. The calculated 1$a''$ single-particle energies $\epsilon_{1a''}(+,Q_0)$ and $\epsilon_{1a''}(-,Q_0)$ are shown in Fig.~\ref{Fig:variation} with and without the self-energy correction.

\begin{table}\vspace*{-1.5ex}
\caption{The charge-transition levels associated with the As vacancy at GaAs(110), given in eV. Values in brackets refer to the constrained symmetric relaxation of the positively charged vacancy. The quasiparticle band gap of 1.55 eV in this work, calculated at the theoretical lattice constant 5.55 \AA, is close to the experimental value 1.52 eV.}\label{Table:GaAs}
\begin{ruledtabular}
\begin{tabular}{lcl}
& $\epsilon^{+/0}$ & $\epsilon^{0/-}$\\
\hline
LDA (this work)           & 0.24 \quad (0.07) & 0.15\\
LDA (Ref.~\cite{zhang96}) & \multicolumn{1}{l}{0.32} & 0.4 \\
LDA (Ref.~\cite{kim96})   & \multicolumn{1}{r}{(0.10)} & 0.24\\
$G_0W_0$ (this work)      & 0.49 \quad (0.32) & 0.60\\
\end{tabular}
\end{ruledtabular}
\end{table}

In Table~\ref{Table:GaAs} we summarize the results for the As vacancy at GaAs(110). Values in brackets refer to the constrained symmetric relaxation of $V_\mathrm{As}^+$ and are included for the purpose of comparison with earlier studies. In contrast to Refs.\ \cite{zhang96,kim96}, which found a stable neutral charge state within a narrow energy window, our own calculation at the level of the LDA indicates $\epsilon^{+/0} > \epsilon^{0/-}$ and hence a direct transition from the positive to the negative charge state, but the small energetic separation is within the uncertainty of the calculation. With the $G_0W_0$ approximation we find a reversed ordering, which implies the existence of a stable neutral charge state, and a slightly increased splitting of the charge-transition levels.

\begin{table}
\caption{The charge-transition levels associated with the P vacancy at InP(110), given in eV. The quasiparticle band gap of 1.52 eV in this work, calculated at the theoretical lattice constant 5.81 \AA, is close to the experimental value 1.42 eV.}\label{Table:InP}
\begin{ruledtabular}
\begin{tabular}{lll}
& $\epsilon^{+/0}$  & $\epsilon^{0/-}$\\
\hline
LDA (this work)              & 0.47         & 0.54 \\
LDA (Ref.~\cite{ebert00})    & 0.52         &      \\
LDA (Ref.~\cite{qian02})     & 0.388        & 0.576\\
$G_0W_0$ (this work)         & 0.82         & 1.09 \\ 
Expt.\ (Ref.~\cite{ebert00}) & 0.75$\pm$0.1 &
\end{tabular}
\end{ruledtabular}
\end{table}

The charge-transition levels for the P vacancy at InP(110) are listed in Table~\ref{Table:InP}. Our LDA results are similar to those reported previously \cite{ebert00,qian02} and well below the experimentally deduced value of 0.75$\pm$0.1 eV \cite{ebert00}. The $G_0W_0$ approximation, on the other hand, yields a value for $\epsilon^{+/0}$ that lies within the experimental error bar.

In conclusion, we have developed a general computational scheme for the optical defect levels and thermodynamic charge-transition levels of point defects in semiconductors. The method is broadly applicable to the bulk as well as to surfaces. It relies on a separation of structural and electronic energy contributions that can be accurately evaluated within DFT and many-body perturbation theory, respectively. In this way the discontinuity of the exchange-correlation potential as well as other shortcomings of the LDA are treated appropriately. Our calculated (+/0) charge-transition level for the P vacancy at InP(110) is in close agreement with the experimental analysis, confirming the accuracy of this method.

We thank J\"org Neugebauer and Philipp Ebert for helpful discussions. This work was funded in part by the EU through the Nanophase Research Training Network (HPRN-CT-2000-00167) and the Nanoquanta Network of Excellence (NMP-4-CT-2004-500198).

\end{document}